\documentclass[aps,twocolumn,prd,superscriptaddress,letterpaper]{revtex4-1}

\usepackage{dcolumn}
\usepackage{amssymb}
\usepackage{mathtools}  %mathtools is better than amsmath
\usepackage[pdftex]{graphicx}
\usepackage{mathptmx}
\usepackage{bm}
\usepackage{xspace}
\usepackage{caption}
\usepackage{verbatim}
\usepackage{subcaption}
\usepackage{tabularx}

\newcommand{\LOb}{\ensuremath{\mathrm{LO}_\mathrm{B}}\xspace}
\newcommand{\HDa}{\ensuremath{\mathrm{HD}_\mathrm{A}}\xspace}
\newcommand{\HDb}{\ensuremath{\mathrm{HD}_\mathrm{B}}\xspace}
\newcommand{\HDab}{\ensuremath{\mathrm{HD}_\mathrm{AB}}\xspace}
\newcommand{\FINESSE}{\textsc{FINESSE}\xspace}
\newcolumntype{Y}{>{\centering\arraybackslash}X}

\begin{document}

\title{Broadband sensitivity enhancement of detuned\\dual-recycled Michelson interferometers with EPR entanglement}

\author{Daniel D. Brown}\email{ddb@star.sr.bham.ac.uk}
\author{Haixing Miao}
\author{Chris Collins}
\author{Conor Mow-Lowry}
\author{Daniel T\"oyr\"a}
\author{Andreas Freise}\email{adf@star.sr.bham.ac.uk}
\affiliation{School of Physics and Astronomy and Institute of Gravitational Wave Astronomy, University of Birmingham, Edgbaston, Birmingham B15 2TT, United Kingdom}

\date{\today}

\begin{abstract}
We demonstrate the applicability of the EPR entanglement squeezing
scheme for enhancing the shot-noise-limited sensitivity of a detuned
dual-recycled Michelson interferometers. In particular, this scheme is
applied to the GEO\,600 interferometer. The effect of losses
throughout the interferometer, arm length asymmetries, and imperfect
separation of the signal and idler beams are considered.
\end{abstract}

\maketitle

\section{Introduction}
Current and future generations of gravitational wave detectors will
inject squeezed light to improve the quantum noise limited regions of their
sensitivity~\cite{PhysRevD.23.1693, schnabel2010quantum}.
Envisaged upgrades of gravitational wave detectors with a squeezed
light source~\cite{Oelker16} will require external \textit{filter cavities}~\cite{Evans2013, Kwee2014}
to provide a broadband reduction in the quantum noise. These filter cavities 
rotate the squeezed state to provide amplitude-squeezing at
low frequencies to reduce radiation pressure fluctuations and 
phase-squeezing at higher frequencies to reduce shot-noise. 
It has recently been proposed that a broadband reduction of quantum
noise in gravitational wave detectors can be achieved using
a pair of squeezed EPR-entangled beams to produce frequency-dependent squeezing~\cite{Ma17}.
This method promises to achieve a frequency-dependent optimisation of the
injected squeezed light fields without the need for an external 
filter cavity. Although suitable filter-cavities can be designed,
the additional cavity adds further complexity to the interferometer.
\textit{EPR-squeezing} offers an attractive solution to this by
harnessing the quantum correlations generated between a pair of EPR
entangled beams~\cite{Reid89, PhysRevA.81.062301, PhysRevA.67.054302, PhysRev.47.777} and 
effectively utilising the interferometer itself as a filter cavity,
thereby achieving a similar response with minimal additional optical
components.

\begin{figure}[b]
  \centering
  \includegraphics[width=0.38\textwidth]{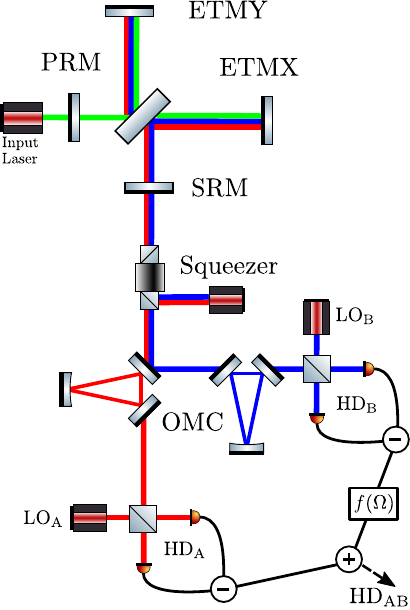}
  \caption{Simplified optical layout of the GEO\,600 detector with
  EPR-squeezing. The main carrier light (green)
  is set to be resonant in the power recycling cavity (PRC). The
  signal (red) and idler (blue) are injected in the dark port and are
  resonant within the signal recycling cavity (SRC). On return from
  the interferometer they are separated and filtered by two
  output mode cleaner cavities (OMCs). }
  \label{fig:epr_layout}
\end{figure}

The GEO\,600 detector in Germany~\cite{Dooley2015} is currently the
only gravitational wave detector to operate for an extended length
of time taking science data using squeezed light to enhance its
sensitivity~\cite{SqueezingNature11, Affeldt14}. GEO is in a prime position to
demonstrate the feasibility of this new technique in the complex
setting of a long baseline interferometer. In this paper we take the
theory suggested previously for the dual-recycled Fabry-Perot Michelson
topology used by LIGO and apply it to GEO, a dual-recycled
Michelson without arm cavities.
The results in this paper were produced using the 
numerical interferometer simulation software
\FINESSE~\cite{Freise04, phd.brown2015}, which allowed
for the correct modelling of quantum noise behaviour 
taking into account the optical losses of
the GEO interferometer.

EPR-squeezing was originally proposed to reduce shot-noise
and radiation pressure noise at the same time. However, the
motivation for its use in GEO would be slightly different:
the sensitivity of GEO\,     is not currently limited by
radiation-pressure noise, as this is masked at low frequencies
by technical noises. However, frequency-dependent squeezing
would be required for GEO to use squeezing effectively in a detuned mode, 
in which the signal recycling cavity (SRC) and thus the peak 
sensitivity of the detector is tuned to a particular offset
frequency. This mode of operation has become of interest again
with new results suggesting that key information about
neutron stars could be obtained from signal frequencies in
the kilohertz region in the ringdown phase after a binary
merger~\cite{Rezzolla2016}. We can show that the EPR squeezing 
technique could be used to operate the GEO detector in such a 
condition with an effective use of squeezed light to reduce 
shot-noise with sufficient bandwidth at the peak sensitivity at 
frequencies around 2\,kHz. We highlight optical design aspects to be
considered for the scheme to be implemented and show how losses
ultimately limit the achievable sensitivity improvements. The
implementation in GEO\,600 would not only allow the improvement
of its sensitivity, it would also serve as a key technology
demonstration for a possible implementation of EPR-squeezing in the
LIGO detectors.

\begin{figure}
  \centering
  \includegraphics[width=0.49\textwidth]{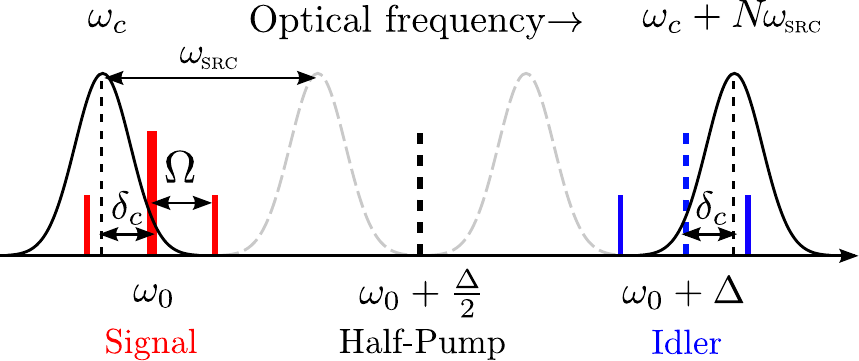}
  \caption{Sketch showing the frequency components and terms used for
    describing the EPR-squeezing scheme. $\omega_c$ is the optical
    frequency of a particular FSR of a cavity. The black lines
    show the signal recycling cavity resonances that the signal and idler
    fields resonate near, the dashed lines represent the $N-1$ FSRs
    between them.}
  \label{fig:epr_spectrum}
\end{figure}

The structure of this article is as follows: in section~\ref{sec:sqz}
we outline the layout of the GEO detector and 
how the EPR-squeezing scheme could be implemented. 
In section~\ref{sec:schnupp}
we model how macroscopic length asymmetries between the arms, the
\textit{Schnupp asymmetry}, must be carefully chosen for the
EPR-squeezing scheme to work. Next, in section~\ref{sec:omc} we look
at the squeezing degradation that occurs at the output-mode-cleaners
due to an imperfect separation of the signal and idler beams. Finally,
in section~\ref{sec:loss} we see how optical losses throughout the
interferometer also affect the sensitivity.

\section{Squeezing with EPR-entanglement in GEO}\label{sec:sqz}

To describe the EPR-squeezing scheme we shall first consider the
simplified layout shown in figure~\ref{fig:epr_layout} along with the
frequency spectrum depicted in figure~\ref{fig:epr_spectrum}.
The pair of EPR entangled beams are generated by an optical parametric
amplifier (OPA). The OPA is pumped at a frequency of
$2\omega_0+\Delta$, generating pairs of entangled sidebands around the
\textit{signal} (red) frequency $\omega_0$, and the \textit{idler}
(blue) frequency $\omega_0 + \Delta$. An incident gravitational wave
will generate a pair of sidebands around the signal carrier at 
frequency $\omega_0\pm\Omega$. No carrier or signal is present around
the idler frequency $\omega_0+\Delta$. 
Unlike the typical squeezing injection which
entangles light at the frequencies $\omega_0 \pm \Omega$, EPR-entanglement
correlates fields at $\omega_0+\Omega$ with fields at $\omega_0+\Delta-\Omega$ and
$\omega_0-\Omega$ with $\omega_0+\Delta+\Omega$. This implies the
quadratures around $\omega_0$ are correlated with those around
$\omega_0+\Delta$, and thus we can reduce the noise at the signal
frequency by making a measurement on the idler---the principal idea
behind conditional squeezing.

The signal and idler beams are injected via a Faraday isolator into the
output path of the interferometer and enter the interferometer through 
the signal recycling mirror (SRM). 
The resonance
condition of the dual-recycled interferometer is set so that the main
carrier light (green in figure~\ref{fig:epr_layout}) destructively
interferes going towards the SRM and constructively back towards the
power recycling mirror (PRM)---this is known as operating at the
\textit{dark fringe}. In this configuration light entering through the
SRM is fully reflected back into this port by the interferometer.
The SRM and the end test
masses, ETMX and ETMY, form the two cavities SRX and SRY. The combination
of both of these is referred to as the \textit{signal recycling
  cavity} (SRC)---a similar argument is made for the PRM, forming
PRX, PRY, and the PRC. Table~\ref{tbl:geo} provides a list of
parameter values used in this work for our GEO600 model.

\begin{table}
\begin{tabularx}{0.45\textwidth}{@{}YY@{}}
 \hline 
 \hline
 \textbf{Parameter} & \textbf{Value} \\
 \hline
 Arm length & 1.2~km \\
 SR length & 1~m \\
 PR length & 1.15~m \\
 \\
 $T_\mathrm{PRM}$ & 900~ppm \\
 $T_\mathrm{SRM}$ & 0.02 \\
 $T_\mathrm{BS}$ & 0.5 \\
 $T_\mathrm{ETM}$ & 0 \\
 \\
 $\omega_{\mathrm{SRC}}$ & $2\pi\cdot125$~kHz \\
 $\omega_{\mathrm{OMC}}$ & $2\pi\cdot435$~MHz \\
 \\
 Input power & 2~W \\
 \hline 
 \hline
\end{tabularx}
\caption{Interferometer parameters used in our GEO600 model}
\label{tbl:geo}
\end{table}

The optical frequency difference between the idler and signal beams,
$\Delta$, must be set such that it is close to an
%chosen so that each beam's frequency is near some
integer number of SRC free spectral ranges (FSR). In particular, with a
SRC detuned by a frequency $\delta_c$, the signal and idler frequency
difference should be
\begin{equation}
\Delta = N \omega_\mathrm{SRC} - 2 \delta_c
\label{eq:sqz_freq_diff}
\end{equation}
(See appendix~\ref{app:cav}) for an interferometer free of any defects.
In this work we study the effect of
EPR-squeezing for an exemplary SRC detuning of 2\,kHz including 
optical losses and asymmetries in the interferometer. The exact value
of the frequency is not important for the arguments made here. It 
was chosen because in principle it allows the improvement of 
quantum noise unimpeded by other technical noises, and is possibly of interest in
the analysis of signals from neutron star mergers.

The detection scheme requires that the signal and idler frequency components  
are spatially separated and measured individually via balanced
homodyne detection. In practice this separation is achieved using a
small cavity such as an output mode cleaner (OMC). This cavity must be
impedance matched for transmission for one of the signal or idler beams and near
anti-resonance for maximal reflection of the other. The signals from
both homodyne detections are then optimally combined
to produce the final output. For this particular GEO\,600
configuration, with negligible radiation pressure effects, the optimal gain
for the signal recombination is frequency independent and depends only
on the amount of squeezing present at the output (see appendix):
\begin{equation}
K_{\rm opt} = \pm \tanh(2r),
\end{equation}
where $r$ is the squeezing factor.
To summarise, four parameters need to be carefully tuned for an
optimal readout: the separation frequency $\Delta$, both local
oscillator (LO) phases of the homodyne readouts, and the gain factor
for the signal recombination.  For the interested reader
a more mathematical description can be found in both the appendices of
this paper and in the supplemental materials of reference~\citep{Ma17}.

\begin{figure}
  \centering
  \includegraphics[width=0.49\textwidth]{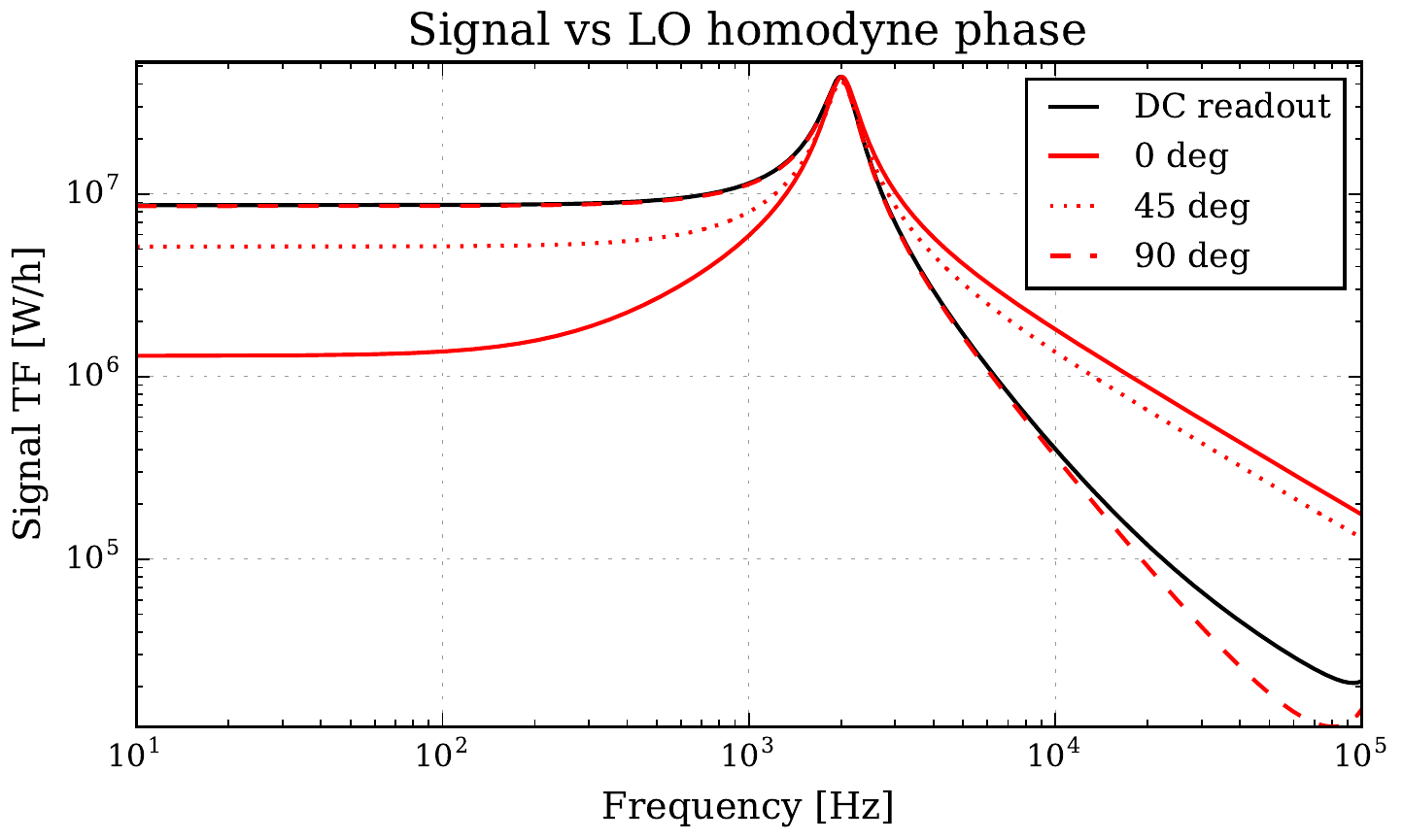}
  \caption{Signal response of GEO\,600 with homodyne detection for
    different homodyne readout angles. This works assumes a 90 deg
    readout which allows for a direct comparison of the new scheme 
    against the current DC readout scheme.}
  \label{fig:geo_homodyne_signal}
\end{figure}

In order to tune the local oscillator phases we start with 
the homodyne detector measuring the
signal beam, in figure~\ref{fig:epr_layout} this is detector \HDa.
The LO phase must be chosen to optimise the detector's susceptibility
to a gravitational wave signal with amplitude $h$, the transfer
function of such a signal to the output \HDa is shown in
figure~\ref{fig:geo_homodyne_signal}. 
The homodyne angle offers a trade-off between high and low frequency
susceptibility. In this work we use the 90\,degrees option as it
provides the best broadband response and
is similar to the DC readout scheme for comparing the EPR scheme
against. The numerical value of this angle is of no importance
for the EPR aspect of this scheme.

\begin{figure}
  \centering
  \includegraphics[width=0.45\textwidth]{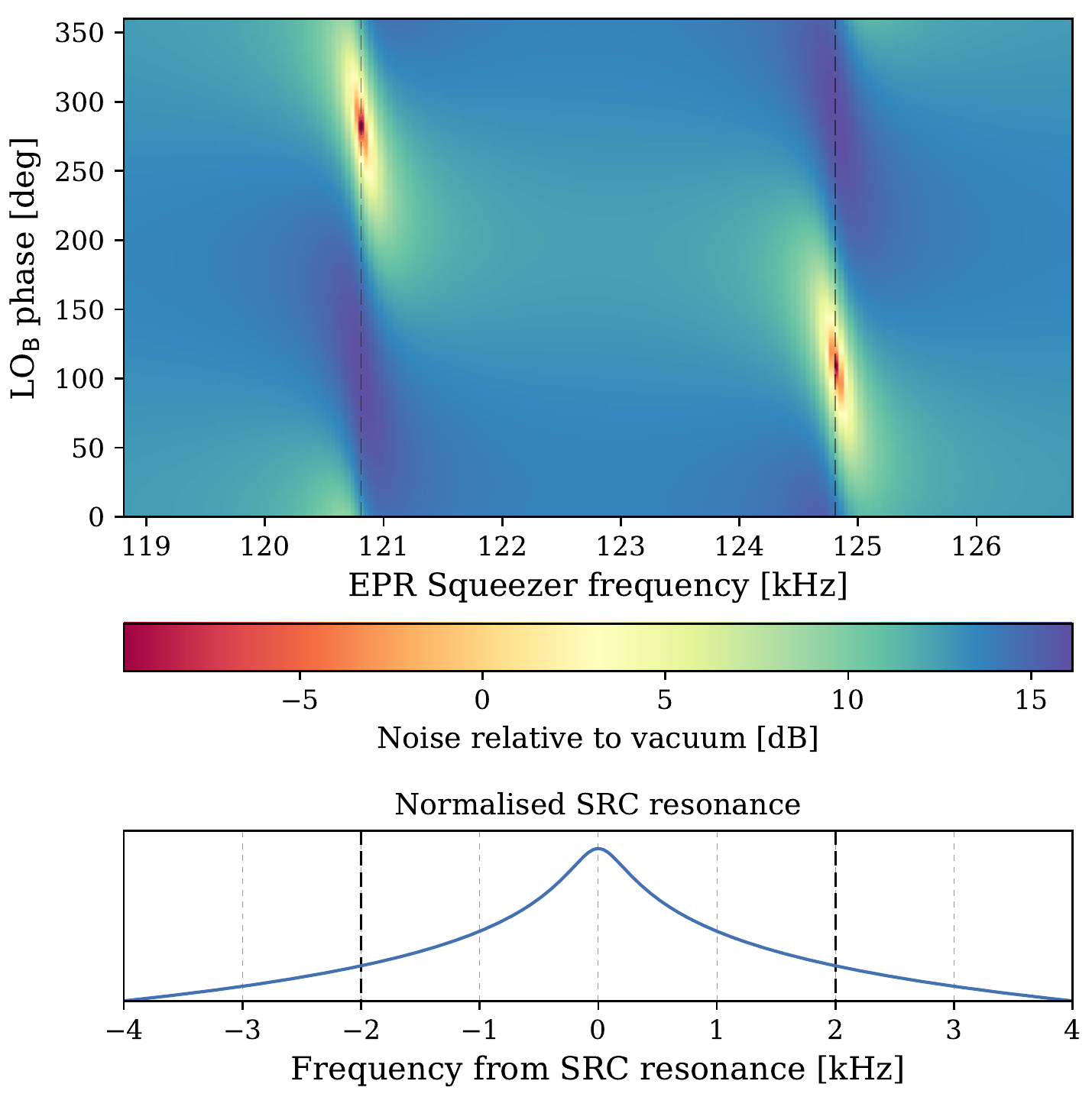}
  \caption{Optimisation for $\Delta$ and the \LOb phase for
    EPR-squeezing.  The interferometer is detuned at
    $\delta_c = 2$\,kHz. The z-axis of the plot shows the noise
    output of \HDab at 2\,kHz. We find two
    potential optimal parameters to choose from, the lower frequency one providing
    the correct broadband noise reduction.  The resulting sensitivity of
    both are shown in figure~\ref{fig:optimised_sensitiviy}.}
  \label{fig:optimise_LOB_sqzf}
\end{figure}

With the required signal homodyne phase being fixed, both $\Delta$ and
the \LOb homodyne phase must then be optimised. The optimal conditions for
these parameters are those that provide the broadest sensitivity
around the detuning frequency. This can be achieved by creating a cost function
describing the squeezing improvement over the desired frequency
range to use with an optimisation routine. However, a
simpler approach was taken here: it is possible to compute
the \HDab output at just the chosen detuning frequency and maximise
the relative noise improvement compared to no squeezing injected. In
this case we find two optimal points around each of the SRCs FSRs
($\approx 125$\,kHz). An example of this optimisation is shown in
figure~\ref{fig:optimise_LOB_sqzf} for a detuning of 2\,kHz. Here 13\,dB
of EPR squeezing is injected and $\approx 10$\,dB of squeezing is
seen~(see appendix~\ref{app:bs} on why a 3\,dB loss is always
  present when using EPR squeezing). At 2\,kHz above and below the
SRC resonance we observe two optimal squeezing conditions. 
The lower optimal squeezer frequency provides the broadband squeezing 
required, as specified by equation~\ref{eq:sqz_freq_diff},
the higher value being the opposite and producing significant
anti-squeezing away from the peak sensitivity. This is analogous to
choosing the correct or incorrect squeezing angle using standard
squeezing injection with DC readout. Also shown here for reference is
the similarly achievable sensitivity when using an equivalent
traditional squeezing input of 10\,dB with the already used DC readout
technique which cannot provide an optimal broadband sensitivity and
is only equivalent to a correctly tuned EPR-squeezing scheme at the
detuning frequency.

\begin{figure}
  \centering
  \includegraphics[width=0.49\textwidth]{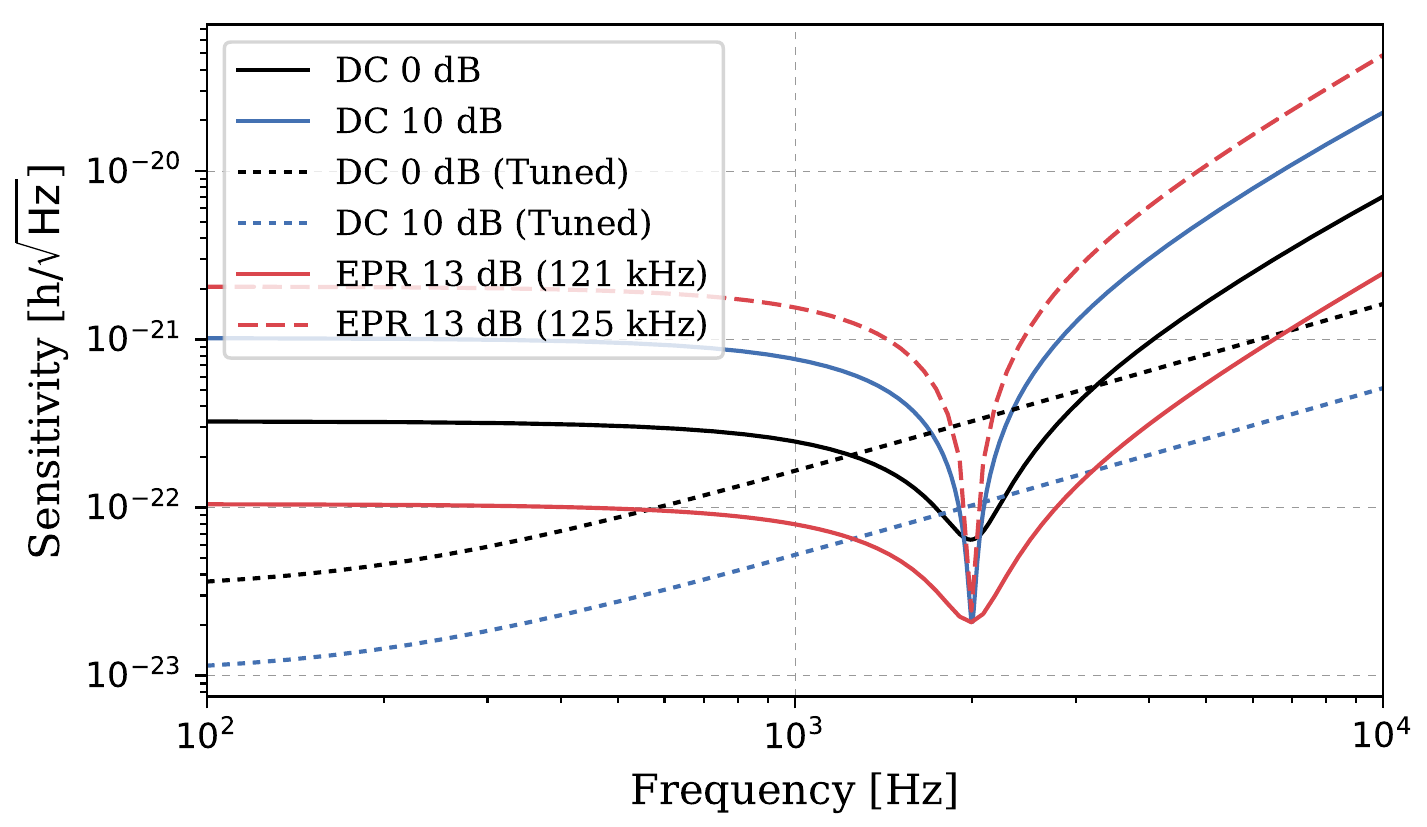}
  \caption{The standard shot-noise limited sensitivity of the lossless GEO
    model (both tuned and detuned) is shown in black. For comparison we show that using 10\,dB of
    squeezing with DC readout results in a slight improvement in the detuned case but
    with significantly narrower bandwidth. The EPR-squeezing has
    two possible parameter choices to optimise the sensitivity at
    $\delta_c$ as shown in figure~\ref{fig:optimise_LOB_sqzf}, however
    only one provides the required broadband improvement.}
  \label{fig:optimised_sensitiviy}
\end{figure}

Using a lossless and symmetric GEO model and the optimal parameters
found in figure~\ref{fig:optimise_LOB_sqzf}, the EPR squeezing for
both optimal parameter choices are compared to DC readout in
figure~\ref{fig:optimised_sensitiviy}. Here we see how an ideal frequency-dependent
squeezing scheme can widen the sensitivity around the detuning frequency, below what
would normally be achievable in the tuned interferometer case with frequency-independent
squeezing. In the following sections we will consider how particular defects
affect the performance of the EPR scheme.

%\db{Add in final plot of EPR-squeezing result using GEO teams best
%guess at current loss values}

\section{Schnupp asymmetry}\label{sec:schnupp}

Radio frequency (RF) optical modulation is employed in gravitational
wave detectors for sensing and control of the position and alignment of
optical components. This requires careful design of cavity lengths
to ensure that particular frequencies resonate within them. 
In addition there is a macroscopic differential length difference between the two
interferometer arms, known as the \textit{Schnupp asymmetry}. This is required so that while
the main carrier light is still near a dark fringe, some RF sidebands
will couple into the output port and thus sense the SRM for control purposes.
These RF modulation frequencies are typically of the order of several MHz,
thus at similar frequencies to that which will be required for the EPR
squeezing.  

\begin{figure}
    \centering
    \begin{subfigure}[b]{0.49\textwidth}
      \centering
      \includegraphics[width=\textwidth]{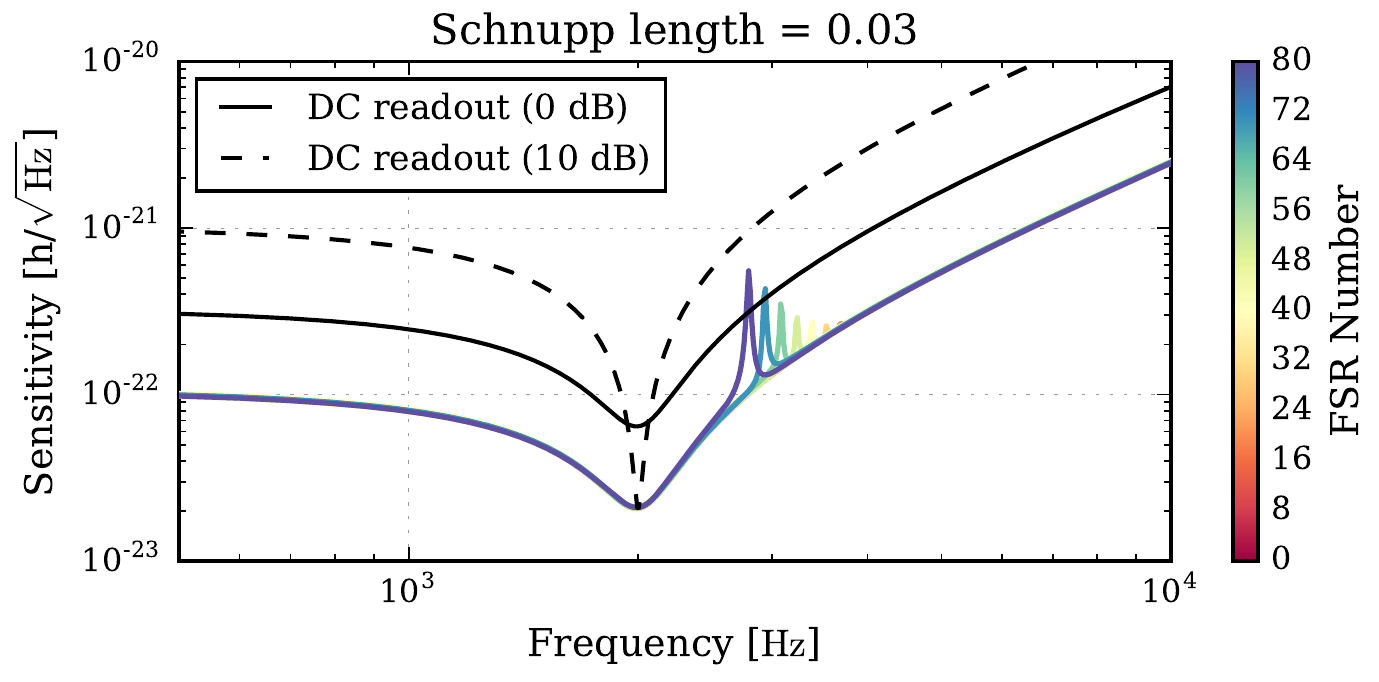}
      \caption{3cm Schnupp asymmetry}
      \label{fig:schnupp_0.03_optimised}
    \end{subfigure}
    \begin{subfigure}[b]{0.49\textwidth}
      \centering
      \includegraphics[width=\textwidth]{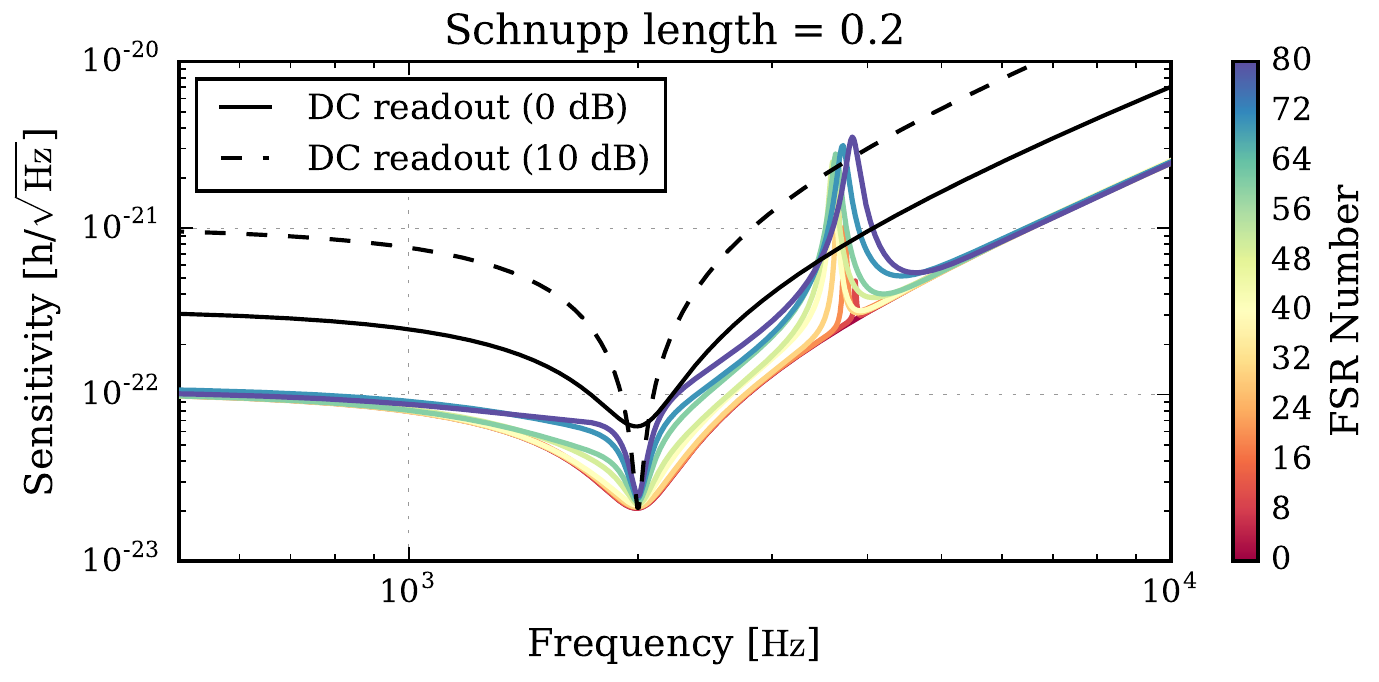}
      \caption{20cm Schnupp asymmetry}
      \label{fig:schnupp_0.2_optimised}
    \end{subfigure}
    \begin{subfigure}[b]{0.49\textwidth}
      \centering
      \includegraphics[width=\textwidth]{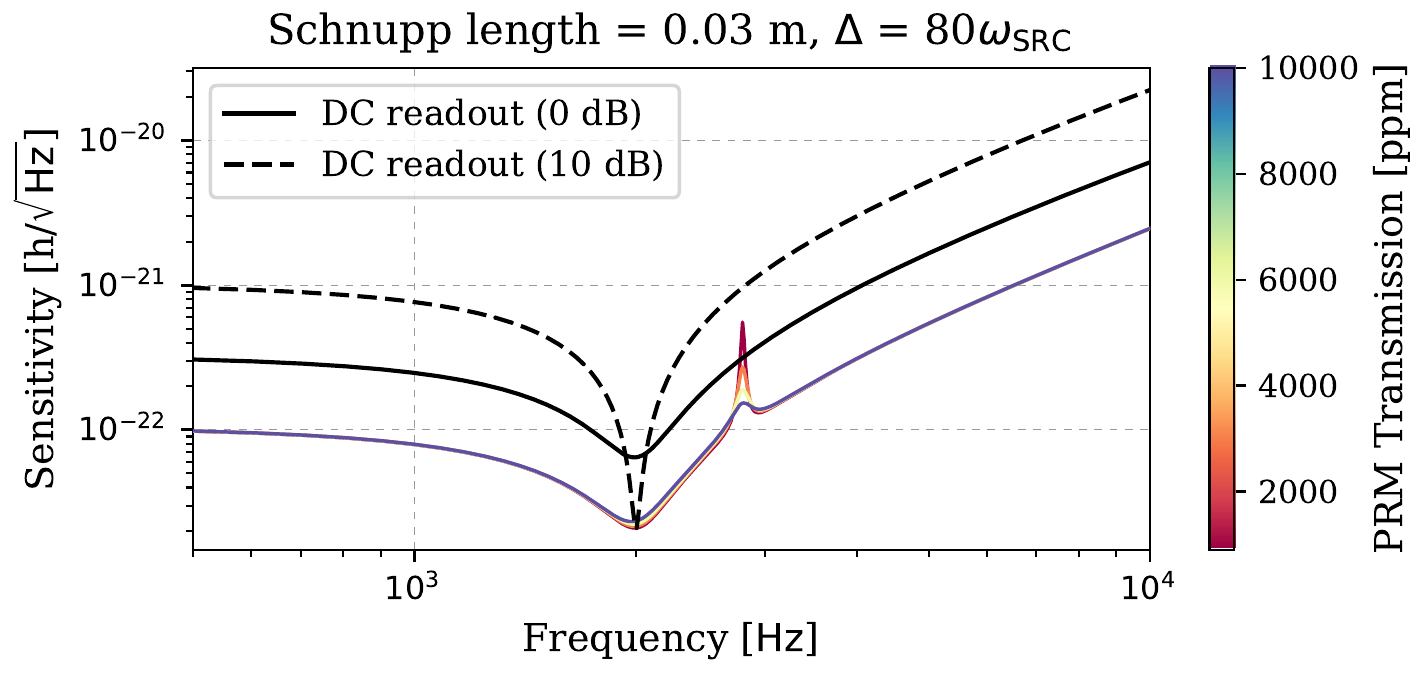}
      \caption{PRM transmission}
      \label{fig:schnupp_prm}
    \end{subfigure}
    \caption{
    The two initial plots illustrate how the choice of the Schnupp asymmetry
    and the corresponding matching of $\Delta$ to $N \omega_\mathrm{SRC}$
    affects the sensitivity. Using a smaller Schnupp asymmetry reduces the additional
    resonance peak and results in less distortion near the peak sensitivity. The final
    plot shows how this additional resonance is affected by the PRM transmission due to
    the coupling of the SRC and PRC due to a Schnupp asymmetry. The optical power in the
    arm cavities were kept constant by adjusting the input power in each case to compare 
    this effect.
    }
    \label{fig:schnupp}
\end{figure}

To implement the EPR squeezing scheme the Schnupp
asymmetry should be chosen to satisfy design requirements for sensing
schemes as well as ensure the higher frequency EPR fields are correctly
detuned from an appropriate SRC resonance. To achieve this, the squeezer frequency should be 
set as an integer number, $M$, of the Schnupp asymmetry FSR, thus both SRX and SRY are on resonance:
$\Delta = M c / (2 L_s)$,
where $L_s$ is Schnupp length difference. 
For technical reasons it is desirable to keep $\Delta$
as low as possible, in the range of 10s of MHz, and at a frequency
that is well reflected by an OMC. The first value, $M=1$, for a
frequency $\Delta=2\pi\cdot10$~MHz we would require $L_s=2.4$~m---which
begins to be unpractically large.

The second option is to use as small a
Schnupp asymmetry as possible instead. Shown in
figure~\ref{fig:schnupp} is how using a Schnupp asymmetry of 3\,cm 
and 20\,cm in a simplified GEO model affects the squeezing. The
typical feature that appears is an additional resonance peak due to
the now different resonant conditions for SRX and SRY and the coupling
this generates with the PRC. Appendix~\ref{app:peak} highlights the behaviour
of these additional peaks in more detail. Figure~\ref{fig:schnupp_prm} demonstrates
how the PRM transmission affects this additional resonance. By lowering the
finesse of the PRC we can reduce this feature.
However, to achieve similar sensitivities the input power would need to be
increased due to the lower PRC recycling gain. 

How the sensitivity is affected by the choice of SRC FSR is also shown in
figures~\ref{fig:schnupp_0.03_optimised} and \ref{fig:schnupp_0.2_optimised}
up to the 80$^\mathrm{th}$ FSR---this being equivalent to $\approx 10$\,MHz. From these figures it is
evident that large Schnupp asymmetries quickly degrade the broadband
sensitivity as higher SRC FSRs are used. Thus a design requirement for
using EPR-squeezing is to use the smallest Schnupp asymmetry possible.
Currently the asymmetry is set to $\approx 5$\,cm in GEO which should still
allow the benefits of the EPR scheme to be experimentally demonstrated.

\section{Separation of signal and idler}\label{sec:omc}
So far we have assumed a perfect separation of signal and idler
in the detection process. Our models used an OMC with a very narrow linewidth to achieve this.
The currently installed  OMC at GEO\,600 has a linewidth of $\approx
1.4$\,MHz, or $\approx 11 \omega_\mathrm{SRC}$, thus the choice
of $\Delta$ must be larger than this.

\begin{figure}
  \centering
  \includegraphics[width=0.49\textwidth]{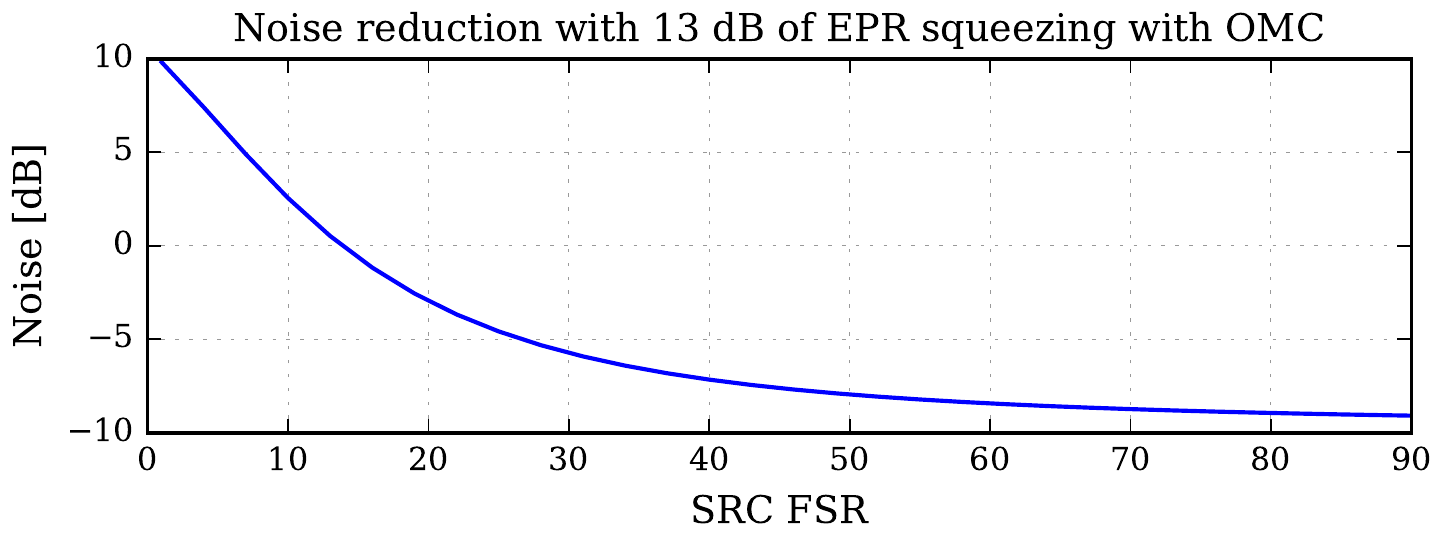}
  \caption{Noise reduction through EPR squeezing as a function of the
  frequency $\Delta$ given in units of free spectral range (FSR) of
  the signal recycling cavity (SRC).
  The current GEO OMC has an linewidth of 1.4\,MHz and FSR of 435\,MHz. The
  SRC FSR is 125\,kHz.  13\,dB of EPR-squeezed light is used to achieve
  a theoretical maximum noise reduction  of 10\,dB.}
  \label{fig:with_OMC}
\end{figure}

Figure~\ref{fig:with_OMC} shows the performance of EPR squeezing if we
replace the perfect OMC with a realistic model of the OMC.
The plot show the best reduction in noise as a function of 
$\Delta$ in units of $\omega_\mathrm{SRC}$. With the
OMC FSR being 435\,MHz, the signal and idler will be ideally separated
at $\approx 217.5$\,MHz $\approx 280\omega_\mathrm{SRC}$. This large separation
is not practical and would require effectively no Schnupp
asymmetry. However, reasonable noise reduction is possible up
to $80\omega_\mathrm{SRC}$, showing a reduction in efficiency of about 
1\,dB compared with an ideal separation of signal and idler.

\section{Susceptibility to optical losses}\label{sec:loss}

Optical losses within an interferometer will degrade any injected
squeezed state by introducing uncorrelated pure vacuum noise. For
current and future squeezing implementations losses will
need to be carefully controlled---for EPR-squeezing the loss requirements
are more strict.
Although in practice losses will occur at each
individual optical component, we can classify the losses into three
categories: a combined input and output loss, 
and internal interferometer symmetric
and asymmetric losses. To depict how these losses affect the
sensitivity, a 2\,kHz detuned GEO model was constructed using a Schnupp
asymmetry of 5\,cm, $\Delta \approx 80$\,SRC~FSRs, a perfectly
separating narrow-band OMC, and 13\,dB of EPR squeezing.
We now compare how three types of loss affects both
EPR-squeezing and an ideal frequency-dependent squeezing source.

\begin{figure}
  \centering
  \includegraphics[width=0.49\textwidth]{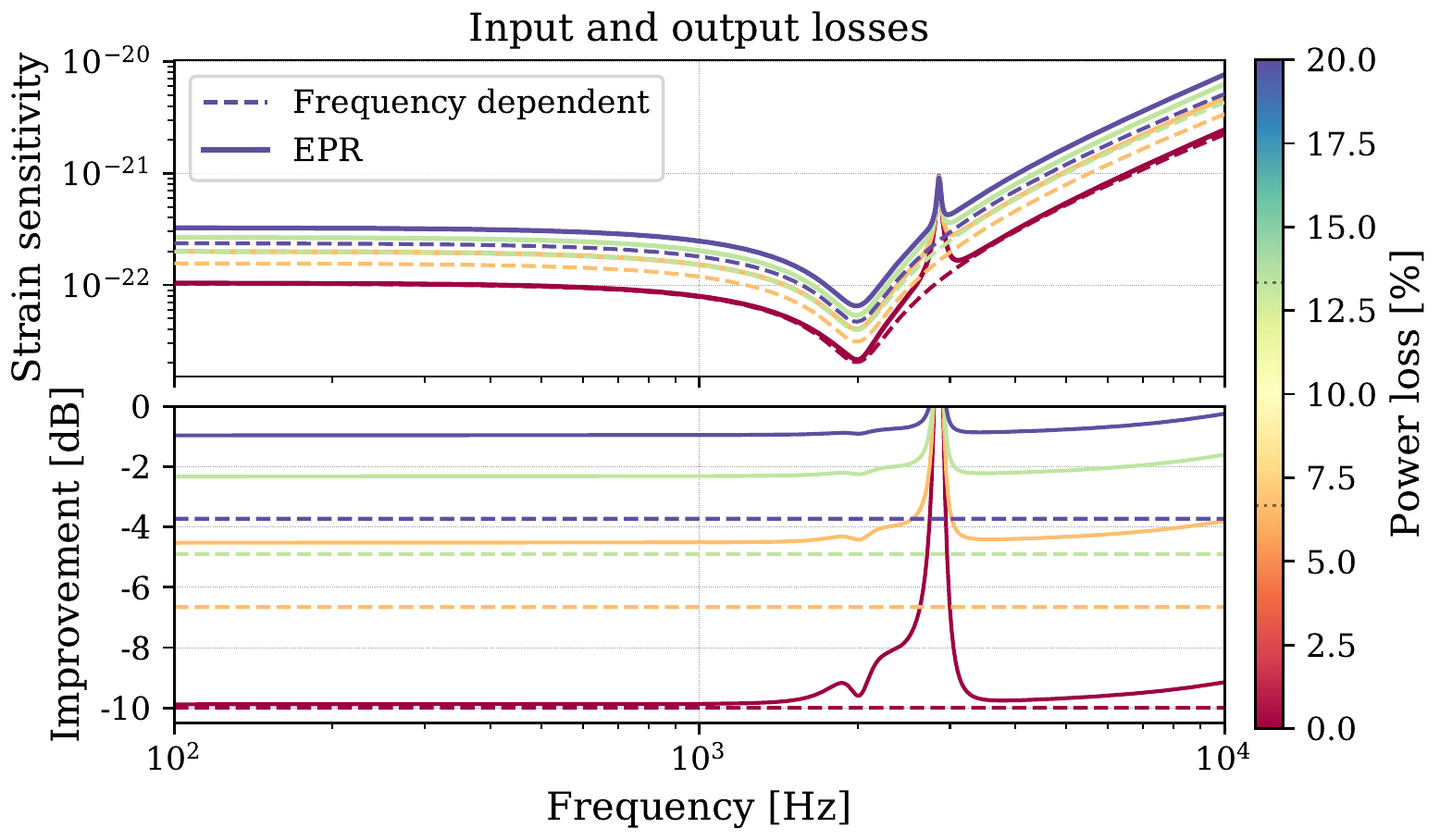}
  \caption{The effect of losses along the input path and output
    path on the detector sensitivity for EPR-squeezing and for an ideal
    frequency-dependant squeezed source. Input and output loss of $X$\% loss refer to an
    $X$\% loss on input plus another $X$\% on output.
    The lower plot shows the overall improvement of the EPR and ideal squeezing
    for identical losses. The peak seen here is due to the chosen Schnupp asymmetry.}
  \label{fig:input_output_loss}
\end{figure}

The combined input and output losses refer to any loss on the
squeezing input path and those on the output path after the SRM up to the
photodiodes. 
Figure~\ref{fig:input_output_loss}  shows how this loss
alters the sensitivity of a detector using an ideal frequency-dependent
squeezed source (dashed) and with EPR-squeezing (solid). Note that a 1\% loss
here means 1\% on input plus another 1\% on output. As expected from Ma`s~work~\cite{Ma17},
EPR-squeezing
is approximately twice as sensitive to optical losses compared to
conventional frequency-dependent squeezing. Without losses
a 10\,dB improvement is seen with some degradation around the
detuning frequency dip and the additional resonance from the Schnupp
asymmetry. A 10\% input and output loss results in a reduction to
around 3\,dB of broadband squeezing. It can also be seen that EPR-squeezing
degrades faster for a given loss value compared to a perfect frequency-dependent
source. 

\begin{figure}
  \centering
  \includegraphics[width=0.49\textwidth]{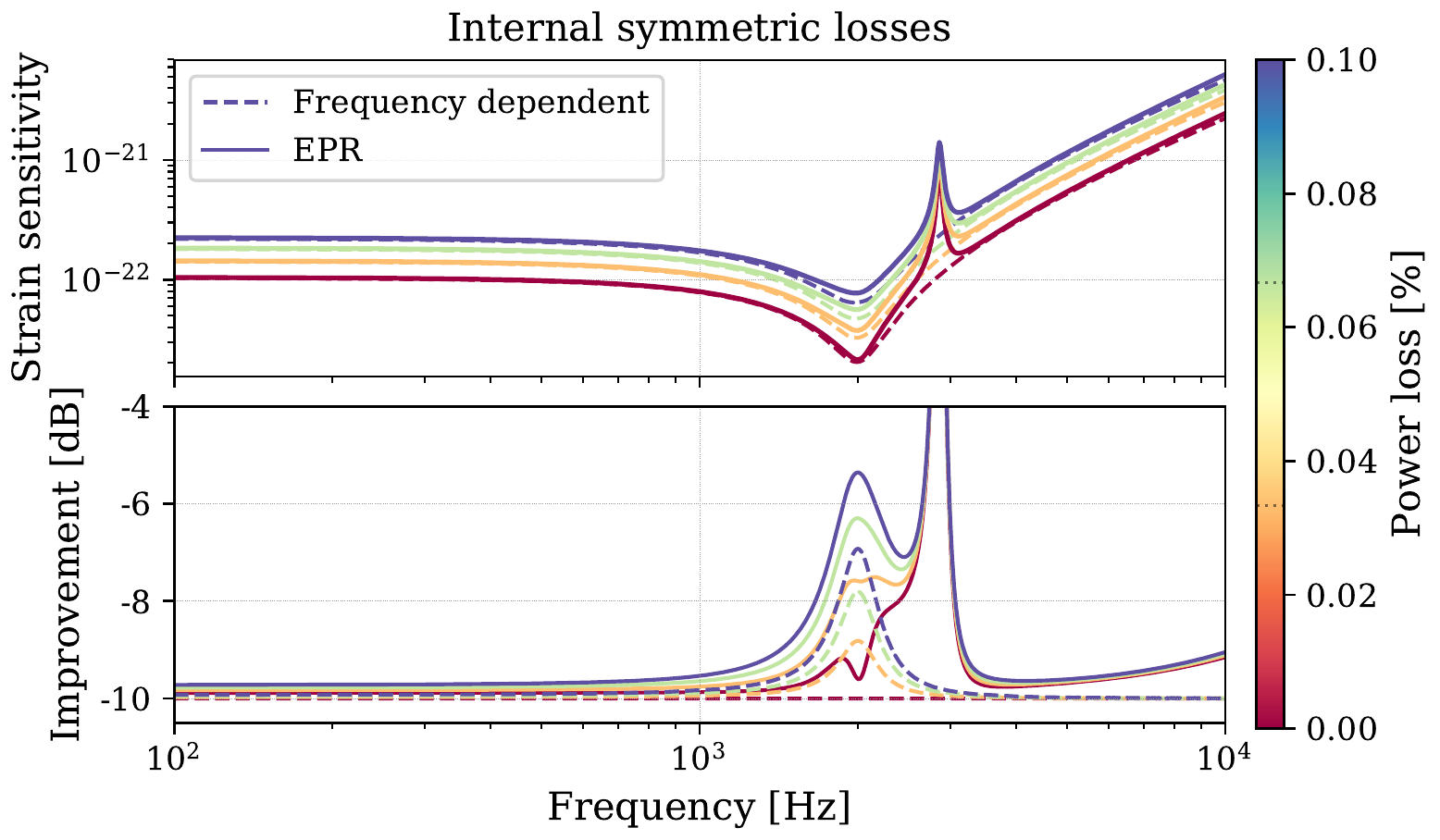}
  \caption{Effect of internal symmetric losses on the sensitivity of the
    interferometer. The bottom plot shows the noise reduction achieved
    by the EPR-squeezing and an ideal frequency-dependent source compared
    to no squeezing in both cases. Significant degradation around the
    detuning frequency are seen in this case.}
  \label{fig:internal_loss}
\end{figure}

\begin{figure}
  \centering
  \includegraphics[width=0.49\textwidth]{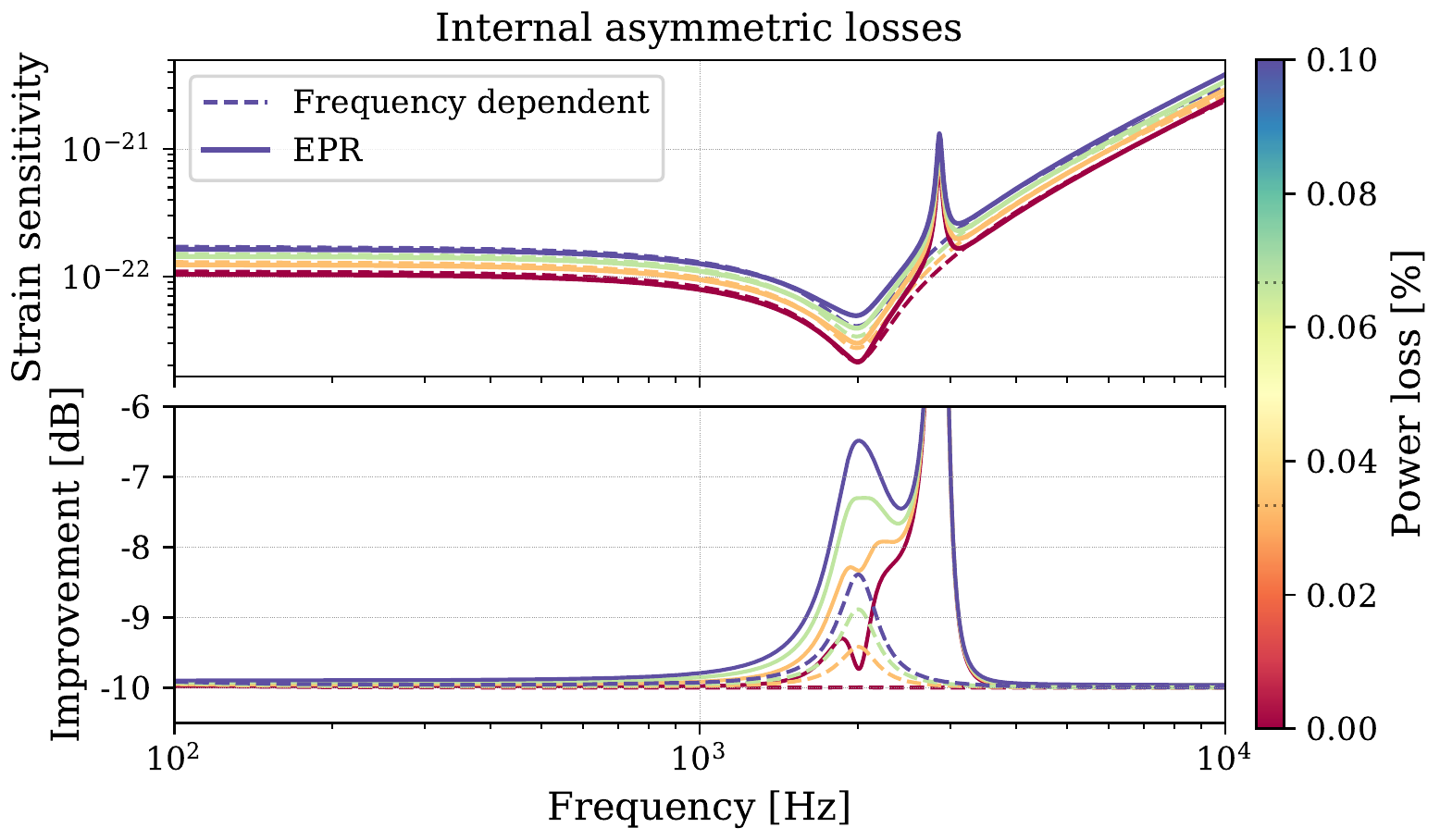}
  \caption{Effect of asymmetric internal loss in the
    interferometer on the sensitivity using 10~dB of ideal frequency-dependent squeezing and
    13\,dB of EPR-squeezing. Asymmetric losses adversely affect the sensitivity around the
    detuning frequency as well as at higher frequencies.}
  \label{fig:asym_internal_loss}
\end{figure}

The internal loss in the  interferometer such as 
clipping from finite optics, surface scattering, or
absorption can be broken down into either
symmetric or asymmetric losses between the two arms.
Figure~\ref{fig:internal_loss} depicts the sensitivity
change due to a range of symmetric losses. 
%Such losses can
%significantly reduce the susceptibility of the detector to a
%gravitational wave. 
At the detuning frequency we see the
squeezing efficiency is affected by losses to a much greater degree due to the
resonance of the signal-recycling cavity multiplying the effect of the loss.
Similarly, the asymmetric losses, as shown in
figure~\ref{fig:asym_internal_loss}, 
also affect the sensitivity predominantly around the detuning frequency.
%Again, in both cases EPR is much more sensitive to large loss values.

It is instructive to compare our result to current estimates of loss
values at GEO600. Internal loss values (symmetric and asymmetric) 
are estimated to be $\sim 0.1\%$. As we have shown, a loss at this level would result
in a substantial reduction in the achievable squeezing around the detuning
frequency. The input/output loss is expected to be $\sim 30\%$ in total, which is
equivalent to a value around 15\% in
figure~\ref{fig:input_output_loss}. Assuming such losses,  13~dB of
EPR-squeezing could still provide a measurable noise reduction and thus be
used for a proof-of-principle demonstration of this technique in a
large-scale laser interferometer. However, for 
operating this scheme in future generations of gravitational wave detectors
losses would need to be reduced significantly.

\section{Conclusion}

We have shown that GEO600 could use the EPR squeezing scheme
to improve the shot-noise limited sensitivity 
with a detuning to 2\,kHz without reducing the detector bandwidth.
We have considered the frequency separation of the signal and idler
beams to be in the MHz range for practical reasons. This in turn
leads to the requirement that the  Schnupp asymmetry must be kept 
as small as possible to suppress additional optical resonances
and to provide the best broadband noise reduction. 
A Schnupp asymmetry of $\approx 3$\,cm would be
sufficient according to our results.
The best value for the Schnupp asymmetry should be based on
a trade-off between this effect and the transmission
of the optical RF sidebands required for controlling the
interferometer.

We have shown how the separation of signal and idler by a
realistic OMC slightly reduces the efficiency of the scheme:
the reduction of effective squeezing 
at around $\Delta \approx 80\omega_\mathrm{SRC} \approx 2\pi\cdot10$\,MHz using the
cavity parameters for the current GEO OMC (excluding its losses) has
been shown to be around 1\,dB. 

Optical losses are limiting the effectiveness of all quantum noise 
reduction techniques. We have demonstrated 
how the sensitivity of GEO\,600 with EPR squeezing is affected by 
losses in the input and output path, and by losses inside the 
interferometer. These results can be used to derive requirements
for potential upgrades for reducing current optical losses to a
level that render the implementation of EPR squeezing feasible.

Overall we found no theoretical design aspects
that would significantly hinder the application of EPR-squeezing in
GEO. The current loss estimates within GEO600 suggest
that EPR-squeezing would offer minimal benefits to the
overall sensitivity of the detector without an addition reduction of
the optical losses. However, GEO600 could provide an important
experimental verification in an active detector of this technique
which is considered an interesting alternative to conventional approach 
of using filter-cavities for detector upgrades and in future
detector designs. 

The authors would like to thank Yanbei Chen and  James Lough for
the idea to apply the EPR scheme in GEO\,600, and  Harald L\"uck,
Hartmut Grote and the GEO\,600 team for their 
support and useful discussions, in particular James Lough and Harald
L\"uck for providing estimates of the optical losses in GEO\,600. This work was
supported by the Science and Technology
Facilities Council Consolidated Grant (number ST/N000633/1) and
H.~Miao is supported by UK Science and Technology Facilities
Council Ernest Rutherford Fellowship (Grant number ST/M005844/11).
D.~T\"oyr\"a is supported by funding from the People
Programme (Marie Curie Actions) of the European Union's Seventh Framework
Programme FP7/2007-2013/ (PEOPLE-2013-ITN) under REA grant agreement number 606176.

\appendix

\section{Peak behaviour: PRC-SRC Schnupp coupling} \label{app:peak}

\begin{figure}
    \centering
    \begin{subfigure}[b]{0.49\textwidth}
      \centering
      \includegraphics[width=\textwidth]{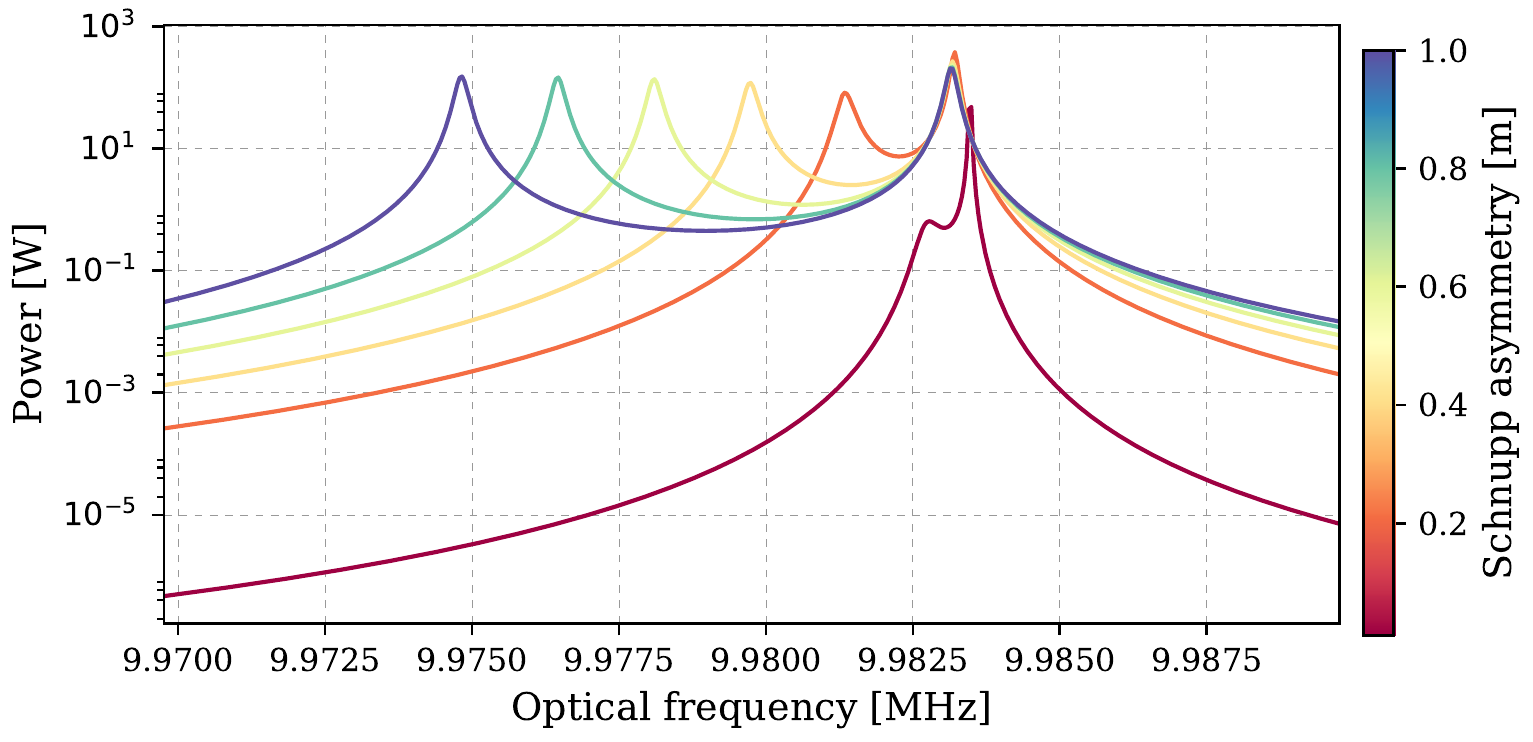}
      \caption{PRC power response}
      \label{fig:schnupp_PRC}
    \end{subfigure}
    \begin{subfigure}[b]{0.49\textwidth}
      \centering
      \includegraphics[width=\textwidth]{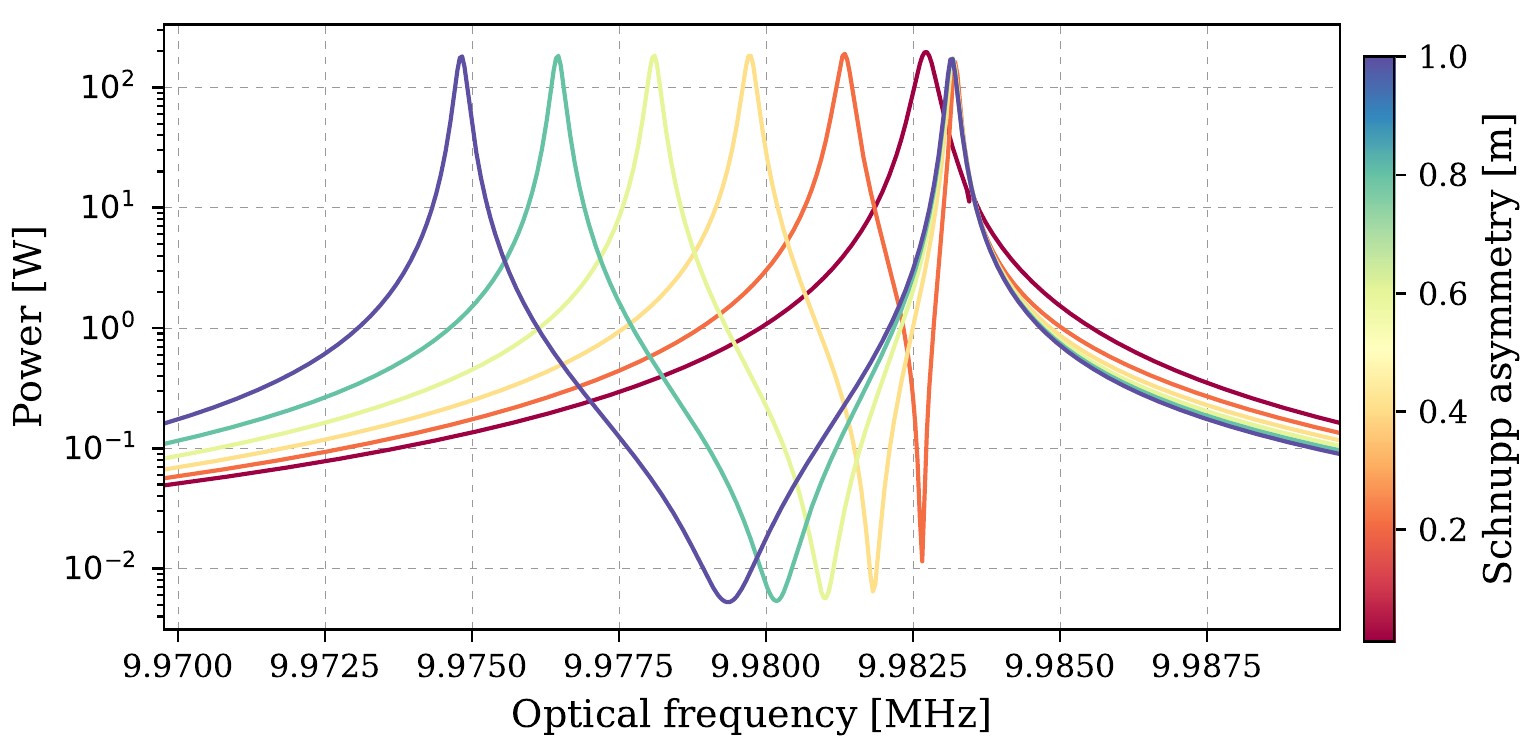}
      \caption{SRC power response}
      \label{fig:schnupp_SRC}
    \end{subfigure}
    \begin{subfigure}[b]{0.49\textwidth}
      \centering
      \includegraphics[width=\textwidth]{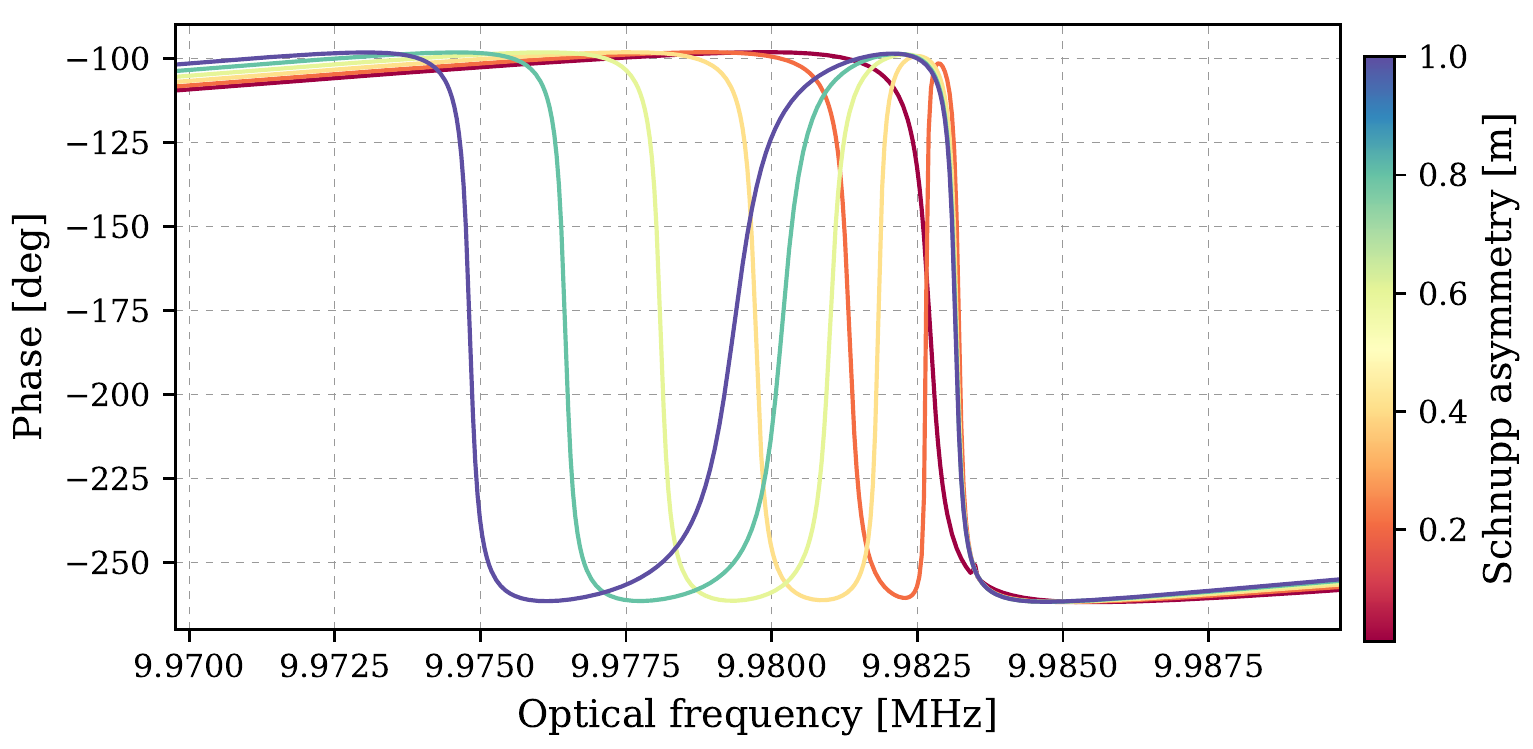}
      \caption{SRC phase response}
      \label{fig:schnupp_SRC_phase}
    \end{subfigure}
    \caption{
    Shown are the powers in the PRC and SRC due to a single frequency optical field injected
    at the dark port. Figure~\ref{fig:schnupp_SRC_phase} shows the phase of the field in the SRC.
    This field's frequency is swept over a similar range to
    what the idler sidebands would be for $\Delta\approx80\omega_\mathrm{SRC}$,
    to visualise how they will react to a weak
    or strongly coupled cavity due to asymmetries. The sharper features seen are from
    the PRC resonance due to its higher finesse. As the asymmetric coupling is
    increased the PRC and SRC become strongly coupled and a split resonance is seen.
    The new resonances also alter the phase of the sidebands in the SRC affecting the correct
    squeezing angle rotation.
    }
    \label{fig:schnupp_coupling}
\end{figure}

By introducing a coupling between the PRC and SRC,\ additional peaks appear in 
the sensitivity spectrum, as shown in figure~\ref{fig:schnupp}. There are two 
regimes to consider: the weak (fig.~\ref{fig:schnupp_0.03_optimised}) and
strong (fig.~\ref{fig:schnupp_0.2_optimised}) coupled cases. When a coupled
cavity becomes strongly coupled an additional resonance is present. We can see
this in figure~\ref{fig:schnupp_coupling}. This shows the power build up in
the PRC-SRC coupled cavity due to a 1~W optical field injected at the dark port
for a range of asymmetries. This is injected at a frequency $\sim 80\omega_\mathrm{SRC}$
to model how the idler fields respond to a coupled system. For the signal
fields this coupling is negligibly small and only a single resonance is seen.

With the idler sidebands seeing a different
optical response from the signal sidebands the optimal sensitivity achievable
is degraded as seen in the previous sections. In figure~\ref{fig:schnupp_coupling}
we see for small Schnupp asymmetries the PRC power is lower and the SRC is a single peak.
As the asymmetries are increased the PRC power is of the order or greater than
that in the SRC and the two become strongly coupled. From here an additional resonance
is visible which further separate in frequency space as the coupling strength increases.

In figure~\ref{fig:schnupp} we see the new resonance beginning to appear around 4~kHz. This is
determined by the choice of $\delta_c$. The idler is is offset from the SRC resonance by $-\delta_c$
and the PRC resonance is $+\delta_c$. Thus at $2\delta_c$ the idler's upper sideband resonates
in the PRC when coupling between the SRC and PRC is present.

In figure~\ref{fig:schnupp} this new resonance drifts shifts in frequency
depending on the asymmetry and chosen $\Delta$. The broadband squeezing is achieved
by correctly rotating the squeezed state, which is determined by the relative phase between
the upper and lower signal and idler sidebands accumulated on reflection from a cavity~(see~\ref{eq:rot_angle}).
When determining the idler's carrier frequency value by optimising $\Delta$,
the lower frequency SRC peak is found to provide the best broadband noise reduction.
Using the lower peak means that the upper idler sideband then interacts with the resonance
conditions that appear.
Figure~\ref{fig:schnupp_SRC_phase} depicts the phase of an optical field reflected from the SRC
over the frequency range of the idler sidebands.
We can see there is a fast change in phase of the sideband around the SRC resonance.
The new resonance, from the strongly coupled cavities, introduces a second phase jump resulting
in an incorrect rotation of the squeezed state at particular frequencies.
This being the reason for the additional peaks in figure~\ref{fig:schnupp}.

\section{Entanglement at a beam splitter}\label{app:bs}

\begin{figure}[htb]
\begin{center}
\includegraphics[width=\columnwidth]{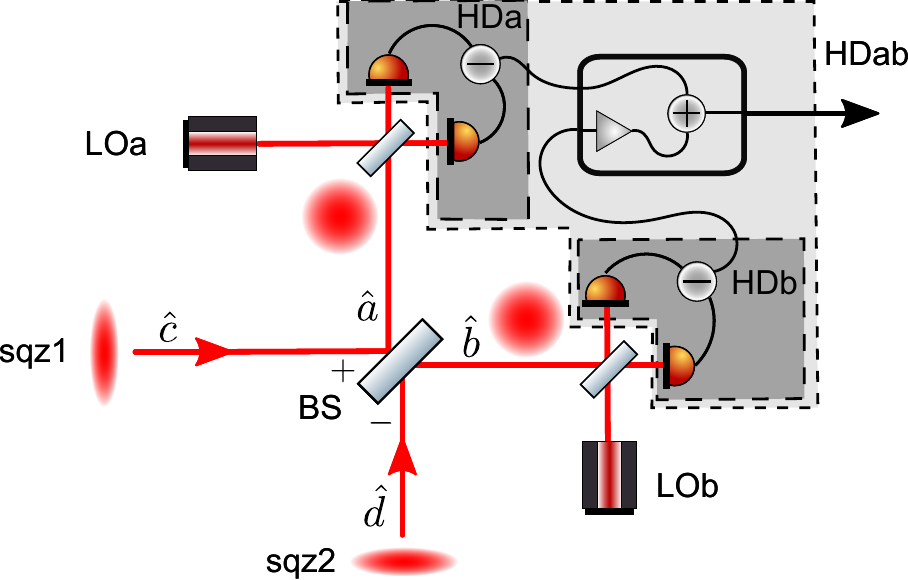}
\caption{Entanglement through overlapping two squeezed fields.}
\label{fig:simpleBS_layout}
\end{center}
\end{figure}
%This appendix outlines the analytics behind the EPR squeezing
%scheme studied in this work. We make the comparison between measuring
%two incident entangled fields on a beamsplitter with the frequency
%separation proposed using a three-mirror cavity.
 
As shown in figure~\ref{fig:simpleBS_layout}, we denote the 
incoming entangled fields from the west port of 
the beamsplitter as $\hat c$ and the south port as 
$\hat d$, and the outgoing field to the 
north port as $\hat a$ and east port as $\hat b$. They satisfy 
the following input-output relation: 
\begin{equation}\label{eq:BS}
\hat a =\frac{1}{\sqrt{2}}(\hat c+\hat d)\,,\quad
\hat b =\frac{1}{\sqrt{2}}(\hat c-\hat d)\,.
\end{equation}
In terms of the amplitude quadrature 
$\hat o_1=(\hat o+\hat o^{\dag})/\sqrt{2}$ 
and phase quadrature $\hat o_2=
(\hat o+\hat o^{\dag})/(\sqrt{2}i)$, the above input-output 
relation can be rewritten as
\begin{equation}\label{eq:BS_quadrature}
\left[\begin{array}{c}
  \hat a_1 \\
  \hat a_2 \\
  \hat b_1 \\
  \hat b_2 
\end{array}\right] 
=\frac{1}{\sqrt{2}}\left[ \begin{array}{cccc}
                       1 & 0 & 1 & 0 \\
                       0 & 1 & 0 & 1 \\
                       1 & 0 & -1 & 0 \\
                       0 & 1 & 0 & -1 
                     \end{array}\right]
\left[\begin{array}{c}
  \hat c_1 \\
  \hat c_2 \\
  \hat d_1 \\
  \hat d_2
\end{array}\right]\,. 
\end{equation}

The covariance matrix of the ingoing field is defined as 
\begin{equation}\label{eq:cov_def}
 {\bf V}_{\rm in}=\langle \psi| \left[\begin{array}{c}
  \hat c_1 \\
  \hat c_2 \\
  \hat d_1 \\
  \hat d_2
\end{array}\right]
[\begin{array}{cccc}
  \hat c_1 & \hat c_2 & \hat d_1 & \hat d_2
\end{array}]
|\psi \rangle\,
\end{equation}
with $|\psi\rangle$ being the quantum state of the optical field. 
Assuming that $\hat c$ is  amplitude squeezed and 
$\hat d$ is phase squeezed (illustrated by the noise ellipse in 
figure~\ref{fig:simpleBS_layout}) we have
\begin{equation}\label{eq:cov_ingoing}
{\bf V}_{\rm in} =
\left[
\begin{array}{cccc}
  e^{-2r} & 0 & 0 & 0 \\
  0 & e^{2 r} & 0 & 0 \\
  0 & 0 & e^{2 r} & 0 \\
  0 & 0 & 0 & e^{-2 r} 
\end{array}
\right]\,, 
\end{equation}
in which $r$ is the squeezing factor. The resulting covariance 
matrix ${\bf V}_{\rm out}$ for the outgoing field, using the 
input-output relation 
Eq.\,\eqref{eq:BS_quadrature}, is then 
\begin{equation}\label{eq:cov_outgoing}
{\bf V}_{\rm out}=\left[
\begin{array}{cccc}
  \cosh2r & 0 & -\sinh2r & 0 \\
  0 & \cosh2r & 0 & \sinh2r \\
 -\sinh2r & 0 & \cosh2r & 0 \\
  0 & \sinh2r& 0 & \cosh2r
\end{array}
\right]\,.
\end{equation}

We can see from this that the amplitude quadrature and phase quadrature for 
either the outgoing field $\hat a$ or $\hat b$ are not correlated, 
which is illustrated schematically by using noise circle in  
figure~\ref{fig:simpleBS_layout}. However, $\hat a$ and
 $\hat b$ are mutually correlated, or equivalently forming
a quantum entanglement, manifested by the 
 nonzero off-diagonal terms in the covariance matrix 
 ${\bf V}_{\rm out}$. It is such a correlation that allows 
 us to reduce the uncertainty (variance) of $\hat a$ by 
 making a measurement on $\hat b$, or vice versa. 
 This is the main principle behind the \textit{conditional 
 squeezing}. 
 
Suppose \HDa measures 
\begin{equation}\label{eq:HDa}
\hat a_{\theta}\equiv \hat a_1 \sin\theta +\hat a_2\cos\theta
\end{equation}
and \HDb measures 
\begin{equation}\label{eq:HDa}
\hat b_{\phi}\equiv \hat b_1 \sin\phi +\hat b_2\cos\phi\,. 
\end{equation}
%The question is what the remaining
%uncertainty (variance) of $\hat a_{\theta}$ conditional 
%on the measurement of $\hat b_{\phi}$, i.e., the conditional 
%variance, which can be derived by using the least mean 
%square estimation. 
We construct the following
estimator for $\hat a_{\theta}$ using the measurement 
outcome of $\hat b_{\phi}$:
\begin{equation}\label{eq:estimator}
\hat a_{\theta}^{\rm est}\equiv K \hat b_{\phi}\,,  
\end{equation}
in which $K$ is some coefficient (filter function). The conditional 
variance of $\hat a_{\theta}$ is defined as
\begin{align}\nonumber
V_{a_{\theta}a_{\theta}}^{\rm cond}&\equiv 
\min_K \langle\psi|( \hat a_{\theta} -
\hat a_{\theta}^{\rm est})^2|\psi \rangle \\ \nonumber
%&= \min_K \langle\psi|( \hat a_{\theta} -
%K\,\hat b_{\phi})^2|\psi \rangle\\\nonumber
& = \min_K [V_{a_{\theta}a_{\theta}}-2K
V_{a_{\theta}b_{\phi}}+K^2 V_{b_{\phi}b_{\phi}}]\\\nonumber
&= \min_K  \left[ 
V_{a_{\theta}a_{\theta}} -\frac{V_{a_{\theta}b_{\phi}}^2}
{V_{b_{\phi}b_{\phi}}}
+V_{b_{\phi}b_{\phi}}
\left(K - \frac{V_{a_{\theta}b_{\phi}}}{V_{b_{\phi}b_{\phi}}}\right)^2
\right]\\
%&= V_{a_{\theta}a_{\theta}} -\frac{V_{a_{\theta}b_{\phi}}^2}
%{V_{b_{\phi}b_{\phi}}}\,. 
\label{eq:Va_cond}
\end{align}
The optimal value for $K$ (Wiener filter) is given by 
\begin{equation}\label{eq:Kopt}
K_{\rm opt} = \frac{V_{a_{\theta}b_{\phi}}}{V_{b_{\phi}b_{\phi}}}\,.
\end{equation}

Given the covariance matrix  ${\bf V}_{\rm out}$ shown in 
Eq.\,\eqref{eq:cov_outgoing}
 for the outgoing field, we have 
% \begin{align}\label{eq:Vxx}
%V_{a_{\theta}a_{\theta}} &= \cosh 2r\,, \\
%V_{a_{\theta}b_{\phi}} & = \cos(\theta+\phi) \sinh 2r\,, \\
%V_{b_{\phi}b_{\phi}} & = \cosh 2r\,. 
% \end{align}
%Plugging these into Eq.\,\eqref{eq:Va_cond} and Eq.\,\eqref{eq:Kopt},
%we obtain
\begin{equation}\label{eq:Va_cond_exp}
V_{a_{\theta}a_{\theta}}^{\rm cond} = \cosh 2r -
\cos^2(\theta+\phi)\sinh2r \tanh 2r\,. 
\end{equation}
and 
\begin{equation}\label{eq:Kopt_exp}
K_{\rm opt} =  \cos(\theta+\phi) \tanh2r\,.
\end{equation}
Therefore, to get the minimum conditional variance for 
$\hat a_{\theta}$, we need to measure the proper 
$\hat b_{\phi}$ such that 
\begin{equation}\label{eq:phi_read}
\phi =-\theta\,,\quad{\rm or}\quad\phi=\pm\pi -\theta\,.
\end{equation}
which yields
\begin{equation}\label{eq:Va_cond_min}
V_{a_{\theta}a_{\theta}}^{\rm cond}\Big|_{\rm min} = 
\cosh 2r - \sinh2r \tanh 2r = \frac{1}{\cosh 2r}\,, 
\end{equation}
and 
\begin{equation}\label{eq:K_opt_min}
K_{\rm opt} = \pm \tanh2r\,, 
\end{equation}
in which the sign depends on the choice of $\phi$ in 
Eq.\,\eqref{eq:phi_read}. This is the optimal gain factor to
use when combining the signal and idler beams. Note, that this is a frequency
independent factor, which is not the case when radiation pressure effects are
dominant.
In particular, given 10dB input squeezing
for both $\hat c$ and $\hat d$, the observed conditional squeezing
is approximately equal to 7dB, i.e., 
\begin{equation}\label{eq:dB}
10 \log_{10}(e^{2r})=10\quad\rightarrow \quad 10 \log_{10}(\cosh2r)\approx 7,
\end{equation}
or that the EPR squeezing scheme results in an automatic 3\,dB loss in squeezing.

\section{Entanglement in the squeezer}\label{app:bs_freq}

\begin{figure}[!b]
\begin{center}
\includegraphics[width=\columnwidth]{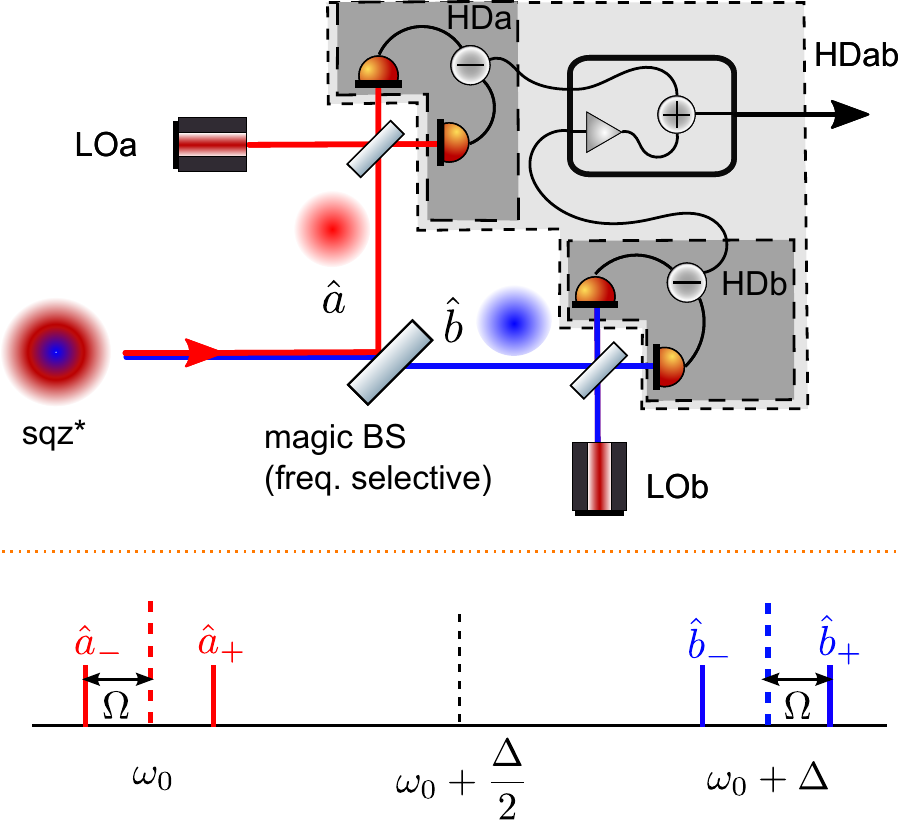}
\caption{Entangled squeezed input and \textit{magic beam splitter}. 
In practice this frequency dependent splitting is achieved through
an optical cavity, reflecting one frequency and transmitting another.}
\label{fig:simple_epr_layout}
\end{center}
\end{figure}

Here we analyse the case of entanglement (correlation) generated 
from a squeezer with a squeezing spectrum over a wide 
frequency range (usually up to 100MHz). The various relevant
frequency for the fields are illustrated in 
figure~\ref{fig:simple_epr_layout}. In particular, 
$\omega_0+\Delta/2$ is half of the pump frequency of the
squeezer. For the usual squeezing injection 
in gravitational-wave detector, this frequency normally 
coincides with the main carrier frequency. However, in the EPR
squeezing scheme, it is intentionally offset 
from the carrier at $\omega_0$ by $\Delta/2$ with
$\Delta$ of the order
of MHz. As a result, the sidebands around $\omega_0$ and those
around $\omega_0+\Delta$ are correlated. Specifically, 
the optical field 
$\hat o(\omega_0 -\Omega)$ is correlated with 
$\hat o(\omega_0+\Delta+\Omega)$, and 
$\hat o(\omega_0 +\Omega)$ is correlated with 
$\hat o(\omega_0+\Delta-\Omega)$. To distinguish 
between the sidebands around $\omega_0$ and those 
around $\omega_0+\Delta$, we introduce
\begin{equation}\label{eq:a_b_sidebands}
\hat a_{\pm}\equiv \hat o(\omega_0\pm \Omega)\,,\quad 
\hat b_{\pm}\equiv \hat o(\omega_0+\Delta\pm \Omega)\,. 
\end{equation}
Given frequency-independent
squeezing source with squeezing factor $r$ and
angle $\theta_s$ ($\theta_s=0$ corresponds to 
phase squeezing), their correlations can be described by using
spectral density, and we have
\begin{align}\label{eq:sqz_spec1}
S_{a_+a_+}&=S_{a_-a_-}
=S_{b_+b_+}=S_{b_-b_-}=\cosh 2r\,, \\
\label{eq:sqz_spec2}
S_{b_-a_+}&= S^*_{a_+b_-} = S_{b_+a_-} =S^*_{a_-b_+} =  e^{2i\theta_s}\sinh2r\,,\\
S_{a_-a_+}&=S_{a_-b_-}=S_{a_+b_+}=S_{b_-b_+} = 0\,, 
\label{eq:sqz_spec3}
\end{align}
where the single-sided spectral density $S_{AB}$ is defined through 
\begin{equation}\label{eq:spec_def}
\frac{1}{2\pi}\langle \psi| \hat A(\Omega)\hat B^{\dag}(\Omega') + 
\hat B^{\dag}(\Omega') \hat A(\Omega) |\psi \rangle \equiv S_{AB}
(\Omega)\,\delta(\Omega-\Omega')\,. 
\end{equation}

\begin{comment}
We follow the usual definition for the amplitude and phase quadratures 
for $\hat a_{\pm}$ and $\hat b_{\pm}$ as
\begin{align}\label{eq:quadrature}
\hat a_1&\equiv 
\frac{1}{\sqrt{2}}[\hat a_++\hat a^{\dag}_-]\,,\quad
\hat a_2\equiv
\frac{1}{\sqrt{2}i}[\hat a_+-\hat a^{\dag}_-]\,,\\ 
\hat b_1&\equiv
\frac{1}{\sqrt{2}}[\hat b_++\hat b^{\dag}_-]\,,\quad 
\hat b_2\equiv
\frac{1}{\sqrt{2}i}[\hat b_+-\hat b^{\dag}_-]\,. 
\end{align}
\end{comment}
With Eqs.\,\eqref{eq:sqz_spec1}, \eqref{eq:sqz_spec2},
and \eqref{eq:sqz_spec3}, 
we can derive the covariance matrix for 
$[\hat a_1\; \hat a_2\; \hat b_1\; \hat b_2]$, in terms of 
spectral density, as 
\begin{widetext}
\begin{equation}\label{eq:cov_mat2}
{\bf S}=\left[\begin{array}{cccc}
        S_{a_1 a_1} &  S_{a_1 a_2} &  S_{a_1 b_1} &  S_{a_1 b_2} \\
        S_{a_2 a_1} & S_{a_2 a_2} & S_{a_2 b_1} & S_{a_2 b_2} \\
        S_{b_1 a_1} &  S_{b_1 a_2} &  S_{b_1 b_1} &  S_{b_1 b_2}  \\
        S_{b_2 a_1} &  S_{b_2 a_2} &  S_{b_2 b_1} &  S_{b_2 b_2} 
      \end{array}
\right]=\left[\begin{array}{cccc}
  \cosh2r & 0 & \cos2\theta_s\sinh2r & \sin2\theta_s\sinh2r \\
  0 & \cosh2r & \sin2\theta_s\sinh2r & -\cos2\theta_s\sinh2r \\
  \cos2\theta_s\sinh2r & \sin2\theta_s\sinh2r & \cosh2r & 0 \\
  \sin2\theta_s\sinh2r & -\cos2\theta_s\sinh2r & 0 & \cosh2r  
\end{array}\right]\,. 
\end{equation}
\end{widetext}
In the special case when $\theta_s=\pi/2$ (amplitude squeezing 
injection), 
the above covariance matrix becomes identical to 
Eq.\,\eqref{eq:cov_outgoing}, i.e., 
\begin{equation}\label{eq:S_0}
{\bf S}|_{\theta_s=\pi/2} = \left[
\begin{array}{cccc}
  \cosh2r & 0 & -\sinh2r & 0 \\
  0 & \cosh2r & 0 & \sinh2r \\
 -\sinh2r & 0 & \cosh2r & 0 \\
  0 & \sinh2r& 0 & \cosh2r
\end{array}
\right]\,. 
\end{equation}
Even though the generation of entanglement
is different from the previous example shown in 
figure~\ref{fig:simpleBS_layout}, the resulting structure 
of entanglement is almost the same, when 
looking at each frequency. Therefore, the reduction 
of uncertainty in $\hat a$, i.e., the conditional
squeezing, by the measurement of $\hat b$ follows 
the same logic as we previously discussed.

\section{Entangled squeezing and frequency independent recombination
  with a simple detuned cavity}\label{app:cav}

\begin{figure}[!b]
\begin{center}
\includegraphics[width=\columnwidth]{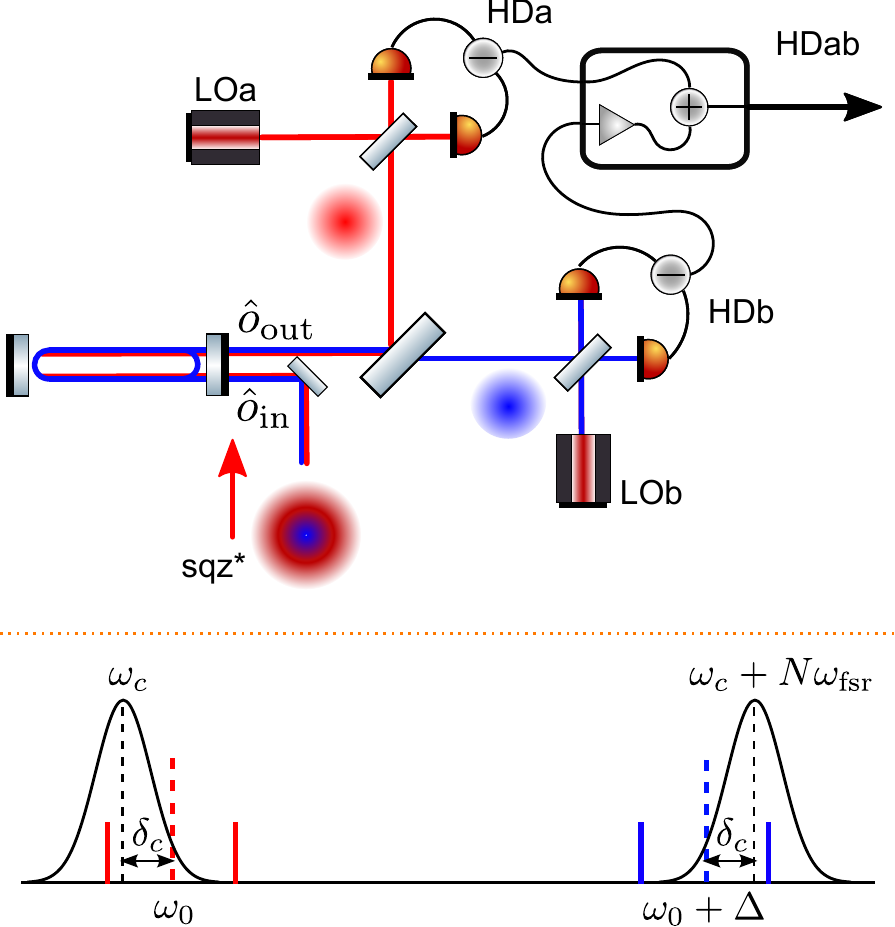}
\caption{Entangled squeezed input reflected of a  cavity gives
  frequency independent squeezing independent squeezing when recombined.}
\label{fig:simple_epr_cavity_layout}
\end{center}
\end{figure}

We now look at the effect of the optical cavity on the 
squeezing field, as shown schematically in 
figure~\ref{fig:simple_epr_cavity_layout}. 
This optical cavity in theory represents the SRC of GEO600.
In the sideband picture, the input-output relation is 
given by 
\begin{equation}\label{eq:cav_io}
\hat o_{\rm out}(\omega) =-\frac{\omega-\omega_{c}-i\gamma_c}
{\omega-\omega_c+i\gamma_c}\,\hat o_{\rm in}(\omega)\,,
\end{equation}
where $\omega_c$ is the cavity resonant frequency, and 
$\gamma_c$ is the cavity bandwidth. 

Take the sideband fields 
$\hat a_{\pm}=\hat o(\omega_0\pm\Omega)$ for example: 
\begin{equation}\label{eq:cav_io_a}
\hat a_{\rm out\pm}=-
\frac{\pm \Omega + \delta_c -i\gamma_c}
{\pm \Omega + \delta_c +i\gamma_c} \hat a_{\rm in\pm}
\equiv e^{i\phi_{\pm}}\hat a_{\rm in\pm}\,. 
\end{equation}
where we have introduced cavity detuning $\delta_c$ and 
sideband phase $\phi_{\pm}$: 
\begin{equation}\label{eq:cav_det}
\delta_c \equiv \omega_0-\omega_c\,,\quad 
\phi_{\pm}\equiv 2\arctan\left(\frac{\pm \Omega+\delta_c}
{\gamma_c}\right)\,.
\end{equation}

In the quadrature picture, the above input-output relation can be 
rewritten as
\begin{equation}\label{eq:cav_io_q}
 \left[\begin{array}{c}
   \hat a_{\rm out 1} \\
   \hat a_{\rm out 2} 
 \end{array}\right] =e^{i(\frac{\phi_+-\phi_-}{2})}
 \left[\begin{array}{cc}
                                                   \cos\frac{\phi_++\phi_-}{2} &  
                                                   -\sin\frac{\phi_++\phi_-}{2} \\
                                                    \sin\frac{\phi_++\phi_-}{2} & 
                                                     \cos\frac{\phi_++\phi_-}{2}
                                                 \end{array}
 \right] \left[\begin{array}{c}
   \hat a_{\rm in 1} \\
   \hat a_{\rm in 2}
 \end{array}\right]\,. 
\end{equation}
Similar relation can also be established between 
$\hat b_{\rm out1, out2}$ and $\hat b_{\rm in1, in2}$. 
As we can see, the quadrature is rotated by a
frequency-dependent angle equal to 
\begin{equation}\label{eq:rot_angle}
\frac{\phi_++\phi_-}{2}=\arctan\left(\frac{\Omega+\delta_c}
{\gamma_c}\right)+\arctan\left(\frac{-\Omega+\delta_c}
{\gamma_c}\right)\,. 
\end{equation}
When the cavity detuning changes sign, the 
rotation angle also changes sign correspondingly, namely, 
\begin{equation}\label{eq:rot_pm}
\frac{\phi_++\phi_-}{2}\Big|_{\delta_c\rightarrow -\delta_c} 
=-\left(\frac{\phi_++\phi_-}{2}\right)\,. 
\end{equation}
Therefore if we arrange the frequency in a way as illustrated in 
figure~\ref{fig:simple_epr_cavity_layout}---$\omega_0$ is 
blue detuned with respect to $\omega_c$ while $\omega_0+\Delta$
is red detuned with respect to $\omega_c+N \omega_{\rm fsr}$ with 
$\omega_{\rm fsr}$ being the free spectral range of the 
cavity, $\hat a_{\rm in1, in2}$ will be rotated opposite to 
$\hat b_{\rm in1, in2}$. From Eq.\,\eqref{eq:phi_read},
this will ensure that the conditional 
squeezing achieves the minimum level at all frequencies, as 
seen in the main results of the paper.

\bibliographystyle{unsrt}

\bibliography{bib}

\end{document}